\documentclass[sigconf]{acmart}

\AtBeginDocument{%
  \providecommand\BibTeX{{%
    \normalfont B\kern-0.5em{\scshape i\kern-0.25em b}\kern-0.8em\TeX}}}

\copyrightyear{2021} 
\acmYear{2021} 
\setcopyright{acmcopyright}\acmConference[ASPDAC '21]{26th Asia and South Pacific Design Automation Conference}{January 18--21, 2021}{Tokyo, Japan}
\acmBooktitle{26th Asia and South Pacific Design Automation Conference (ASPDAC '21), January 18--21, 2021, Tokyo, Japan}
\acmPrice{15.00}
\acmDOI{10.1145/3394885.3431529}
\acmISBN{978-1-4503-7999-1/21/01}

\usepackage[redeflists]{IEEEtrantools}
\usepackage{subcaption}
\usepackage{multirow}
\usepackage{multicol}
\usepackage{flushend}
\usepackage{balance}
\usepackage{microtype}
\usepackage{tabu}
\usepackage{color, colortbl}
\newcommand{\thickhline}{%
    \noalign {\ifnum 0=`}\fi \hrule height 1pt
    \futurelet \reserved@a \@xhline
}
\definecolor{myblack}{rgb}{0,0,0}

\newcommand{\tech}{\text{{HEBE}}}{}
\newcommand{\ineq}[1]{\footnotesize$#1$\normalsize}{}

\begin{document}
\bstctlcite{IEEEexample:BSTcontrol}
\fancyhead{}
\title{Aging-Aware Request Scheduling for Non-Volatile Main Memory}

\author{Shihao Song}
\affiliation{\institution{Drexel University}\country{USA}}

\author{Anup Das}
\affiliation{\institution{Drexel University}\country{USA}}

\author{Onur Mutlu}
\affiliation{\institution{ETH Z{\"u}rich}\country{Switzerland}}

\author{Nagarajan Kandasamy}
\affiliation{\institution{Drexel University}\country{USA}}

\begin{abstract}

    Modern computing systems are embracing non-volatile memory (NVM) to implement high-capacity and low-cost main memory. Elevated operating voltages of NVM accelerate the aging of CMOS transistors in the peripheral circuitry of each memory bank. Aggressive device scaling increases power density and temperature, which further accelerates aging, challenging the reliable operation of NVM-based main memory. We propose \tech{}, an architectural technique to mitigate the circuit aging-related problems of NVM-based main memory. \tech{} is built on three contributions. First, we propose a new analytical model that can dynamically track the aging in the peripheral circuitry of each memory bank based on the bank's utilization. Second, we develop an intelligent memory request scheduler that exploits this aging model at run time to de-stress the peripheral circuitry of a memory bank only when its aging exceeds a critical threshold. Third, we introduce an isolation transistor to decouple parts of a peripheral circuit operating at different voltages, allowing the decoupled logic blocks to undergo long-latency de-stress operations independently and off the critical path of memory read and write accesses, improving performance. 
    We evaluate \tech{} with workloads from the SPEC CPU2017 Benchmark suite.
    Our results show that \tech{}
    significantly improves both performance and lifetime of NVM-based main memory.
\end{abstract}

\maketitle

\section{Introduction}
\label{sec:introduction}
DRAM has been the technology choice for implementing main memory due to its relatively low latency and low cost. 
However, DRAM is a fundamental performance and energy bottleneck in almost all computing systems, and it is experiencing significant technology scaling challenges~\cite{KimISCAA14,kim2020revisiting,mandelman2002challenges,mutlu2013memory,mutlu2017rowhammer,mutlu2019rowhammer,das2018vrl}.
Recently, DRAM-compatible, yet more technology-scalable alternative non-volatile memory (NVM) technologies such as Phase-Change Memory (PCM), are being explored~\cite{LeeISCA2009,lee2010phase,LeeMicro10,yoon2014efficient,mutlu2015research,qureshi2010improving,QureshiISCA09,qureshi2011pay,palp,kultursay2013evaluating,mneme,datacon,wong2010phase,meza2013case}.\footnote{NVMs are also used for synaptic storage in neuromorphic computing~\cite{Burr2017,mallik2017design,spinemap,esl20,dfsynthesizer}.}

Compared to DRAM, NVM requires higher voltages to read and program memory cells. 
We investigate the internal architecture of the peripheral circuitry of each memory bank and find that such circuitry consists of transistors built using CMOS and FinFET~\cite{hisamoto2000finfet}.
When operated at high voltage and temperature, over time the transistor's parameters can strongly drift from their nominal values. This is called \emph{aging}. {In fact, in scaled technology nodes, aging happens even under nominal conditions from the very start of device use.} The most important breakdown mechanism is the Bias Temperature Instability (BTI)~\cite{weckx2014non,kraak2019parametric}. {Strongly depending on the workload, BTI is highly variable and largely reversible under nominal conditions upon removal of the stress voltage. 
However, if the peripheral circuitry is used continuously for long durations at elevated operating conditions, the BTI induced parameter drifts in peripheral circuitry cannot be reversed~\cite{6531944}, leading to permanent functional degradation and hardware faults.}

{As process technology scales down to smaller dimensions due to NVM's CMOS-compatible scaling~\cite{XiongIEDM16}, aging issues are expected to get exacerbated
due to the increase in 
the electric field and power density, 
which 
leads to higher chip temperatures and, consequently, the acceleration of BTI.} 
Current methods for improving aging are overly conservative, since they estimate transistor aging in a peripheral circuitry \emph{statically}, assuming worst-case operating conditions~\cite{jiang2014low}. Based on such worst-case estimates, these methods de-stress each peripheral circuitry periodically at a fixed interval, without tracking its actual aging. Therefore, these methods significantly and unnecessarily constrain performance.

Our \textbf{goal} is to design a dynamic policy to track the aging in the peripheral circuitry of each memory bank based on the operating voltages needed to serve read and write requests from the bank,
and dynamically schedule its de-stress operation only when its aging exceeds a critical threshold.
Our architectural approach to mitigate aging in NVM, called \tech{},\footnote{In Greek mythology, Hebe (pronounced hee.bee) is the goddess of youth~\cite{wiki:hebe}.} is built on three contributions.

\textbf{Contribution 1.} We develop a new, {accurate} analytical model to estimate transistor aging in peripheral circuitry of each memory bank. Our model {dynamically} tracks aging in response to a memory controller's request scheduling decisions such as serving a read (which requires \ineq{2.85V}) vs. a write (which requires \ineq{3.7V}).
To use this model at {run time}, we leverage the {associative property} of our analytical formulation, a direct reflection of the underlying physical failure mechanism, allowing us to express aging in terms of offline-computed \emph{unit aging} parameters (described in Section~\ref{sec:aging_formulation}). {Our memory controller uses these parameters to estimate aging in a peripheral circuitry based on the number of read and write requests that are served via the circuitry.} 

\textbf{Contribution 2.} We develop a new, {intelligent} memory request scheduler that 
prioritizes
requests to banks whose peripheral circuits are currently 
active but not serving any memory request, over other requests, including the long-{outstanding} ones.
This {straightforward} and {greedy} policy is controlled in two ways. 
{First, the memory controller uses our new aging model to 
track the aging of a peripheral circuitry, de-stressing the circuitry only when its aging exceeds a critical threshold.
Second, the memory controller uses thresholding to avoid starvation of memory requests.}

\textbf{Contribution 3.} We introduce an isolation transistor in each peripheral circuitry to decouple 
its logic blocks operating at different supply voltages during read and write accesses
(see Fig.~\ref{fig:peripheral_circuit}). The decoupled architecture allows these logic blocks 
to be de-stressed based on their respective aging levels.
Our request scheduler exploits this decoupled architecture to 
schedule the long-latency de-stress operations off the critical path of accesses, reducing bank occupancy and improving performance.

We evaluate \tech{} with workloads from the SPEC CPU2017 Benchmark suite~\cite{bucek2018spec}.
    Our results show that \tech{}
    significantly improves both performance and lifetime of NVM-based main memory.

\section{Background}
\label{sec:background}
\begin{figure}[h!]
	\centering
	\vspace{-10pt}
	\centerline{\includegraphics[width=0.99\columnwidth]{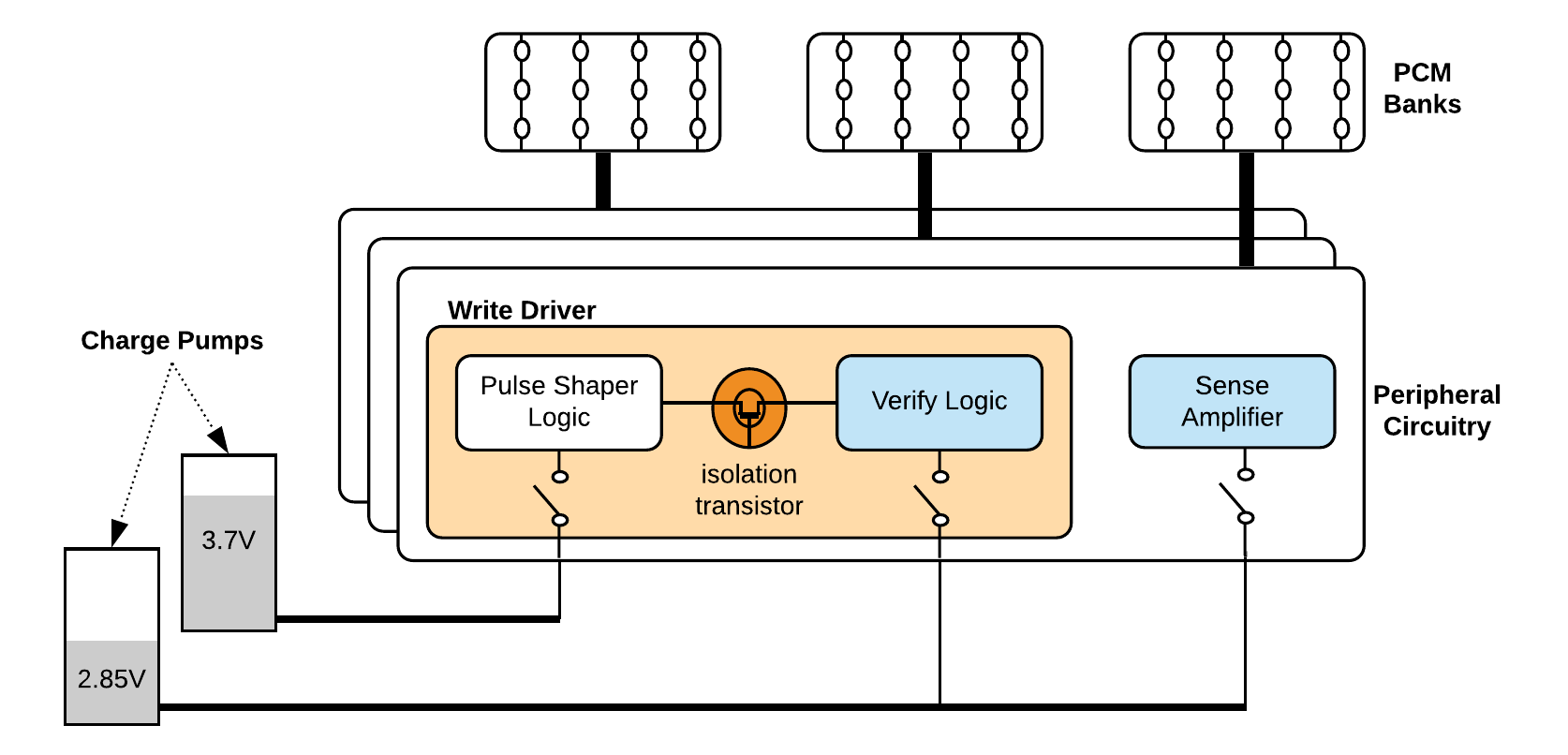}}
	\vspace{-10pt}
	\caption{Architecture of NVM peripheral circuitry \cite{palp}.}
	\vspace{-10pt}
	\label{fig:peripheral_circuit}
\end{figure}

NVM, like DRAM~\cite{KimISCA12,seshadri2019dram,das2018vrl,lee2013tiered}, is organized hierarchically~\cite{pcm_book,palp,mneme,datacon,meza2012case,meza2018evaluating,yoon2014efficient}. For example, a 128GB NVM can have 2 {channels}, 1 rank/channel and 8 banks/rank. A bank can have {64} {partitions}~\cite{palp}.
Each bit in NVM is represented by the resistance of an NVM cell: low resistance is logic `1' and high resistance is logic `0'.
An NVM cell is read and programmed by driving current through it using per bank peripheral circuitry (see Fig.~\ref{fig:peripheral_circuit}). {Peripheral circuitry consists of sense amplifiers (SA) to read and write drivers (WD) to program.} 
WD consists of the {write pulse shaper (PS) logic}, which generates the current pulses necessary for SET and RESET operations, and the {verify (VF) logic}, which verifies the correctness of these operations~\cite{frulio2016adaptive}. 

{In addition to the regular read and write mode of operations, peripheral circuitry can also be in 1) \textit{idle} mode, where it does not serve any request, and 2) \textit{de-stress} mode, where it is powered down.} 
Table~\ref{tab:peripheral_voltages} reports the operating voltages of the three logic blocks in a peripheral circuitry during read, write, idle, and de-stress operations~\cite{pcm_book}.
Voltages higher than the nominal \ineq{1.2V} supply are generated using the two on-chip charge pumps shown in Figure~\ref{fig:peripheral_circuit}. These high voltages induce aging of the transistors in the peripheral circuitry logic blocks. We focus on BTI failures.

\begin{figure}[h!]
 	\centering
    \vspace{-10pt}
 	\centerline{\includegraphics[width=0.6\columnwidth]{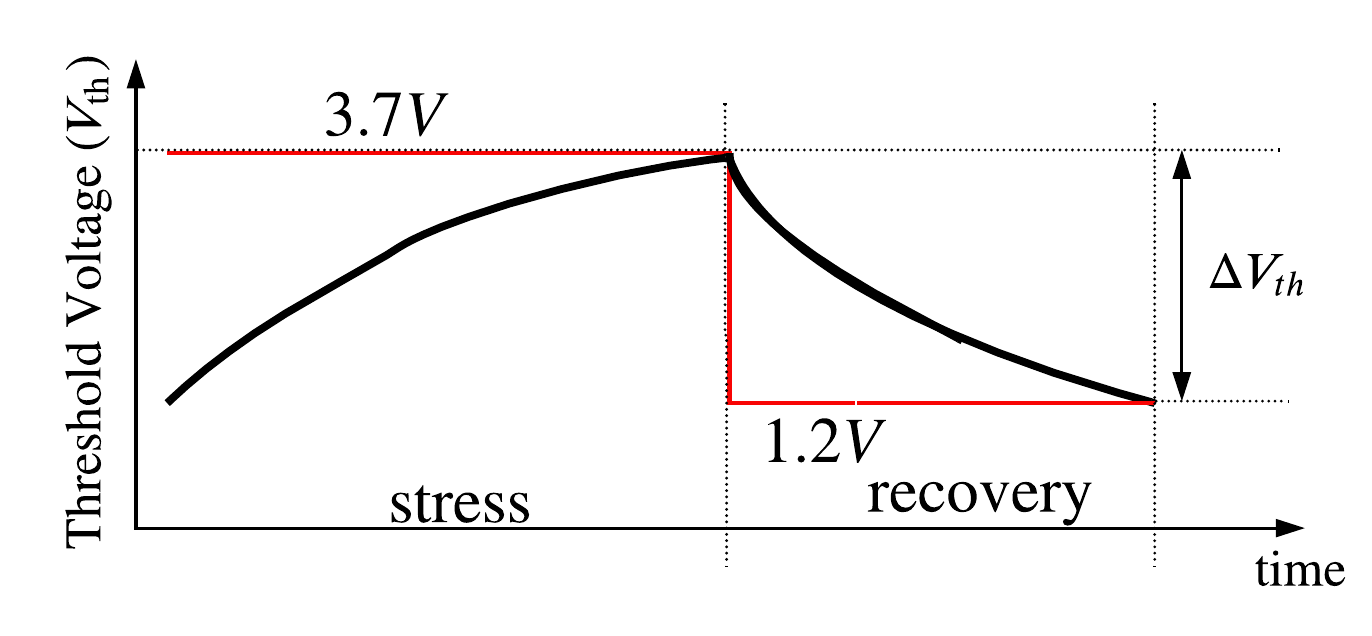}}
 	\vspace{-10pt}
 	\caption{Threshold voltage (\ineq{V_\text{th}}) shift due to BTI.}
    \vspace{-10pt}
 	\label{fig:nbti_demo}
\end{figure}

BTI is a failure mechanism in a transistor, where positive charge is trapped in the oxide-semiconductor boundary underneath the gate~\cite{gao2017nbti}.
BTI manifests as 1) decrease in drain current and transconductance, and 2) increase in off current and threshold voltage \ineq{V_\text{th}}. 
Figure~\ref{fig:nbti_demo} illustrates the stress and recovery of the threshold voltage of a transistor on application of a high (\ineq{V_\text{read}/V_\text{write}}) and a low (\ineq{V_\text{de-stress}}) voltage. We observe that both stress and recovery depends on the time of exposure to the corresponding voltage level. This implies that when peripheral circuitry is de-stressed, the BTI aging of its transistors partially recovers from stress.
To compute the overhead due to de-stress operations, we assume that the memory controller
issues a de-stress command to a memory bank once every \ineq{tDSI}, the \textit{de-stress interval}. Each de-stress operation completes within a
time interval \ineq{tDSC}, the \textit{de-stress cycle time}. 
Hence, the performance overhead (i.e., data throughput loss) due to periodic de-stress is 
\begin{equation}
    \label{eq:destress_overhead}
    \scriptsize \text{de-stress overhead} = tDSC/tDSI.
\end{equation}
The overhead due to periodic de-stress 
(as implemented in conservative approaches such as \cite{jiang2014low}) 
is significant in current NVM devices, and it is expected to become even more performance-critical in the future as NVM chip capacity increases~\cite{pcm_book}. 

\vspace{-5pt}
\begin{table}[h!]
\setlength{\tabcolsep}{7pt}
\renewcommand{\arraystretch}{1.0}
\centering
{\fontsize{8}{10}\selectfont
\begin{tabu}{c c c c}
    \tabucline[2pt]{-}
    \multirow{3}{*}{\textbf{Operating mode}} & \multicolumn{3}{c}{\textbf{Operating voltage}} \\ \cline{2-4}
    & \textbf{pulse shaper} & \textbf{verify} & \textbf{sense amplifier} \\
    & \textbf{(PS)} & \textbf{(VF)} & \textbf{(SA)}\\
    \hline
    Read & 1.2V & 1.2V & \textcolor{red}{2.85V}\\
    Write (program) & \textcolor{red}{3.7V} & \textcolor{red}{2.85V} & 1.2V\\
    Idle & 1.2V & 1.2V & 1.2V\\
    De-stress & \textcolor{blue}{$<V_\text{th}$} & \textcolor{blue}{$<V_\text{th}$} & \textcolor{blue}{$<V_\text{th}$}\\
    \tabucline[2pt]{-}
\end{tabu}
}
\vspace{5pt}
\caption{Operating voltage of the three logic blocks in peripheral circuitry during read, write, idle, and de-stress~\cite{pcm_book}. {The threshold voltage (\ineq{V_{th}}) of a CMOS transistor is between 0.7V and 1V at scaled nodes.}}
\label{tab:peripheral_voltages}
\end{table}
\vspace{-20pt}

\begin{figure}[h!]
 	\centering
    \vspace{-5pt}
 	\centerline{\includegraphics[width=0.99\columnwidth]{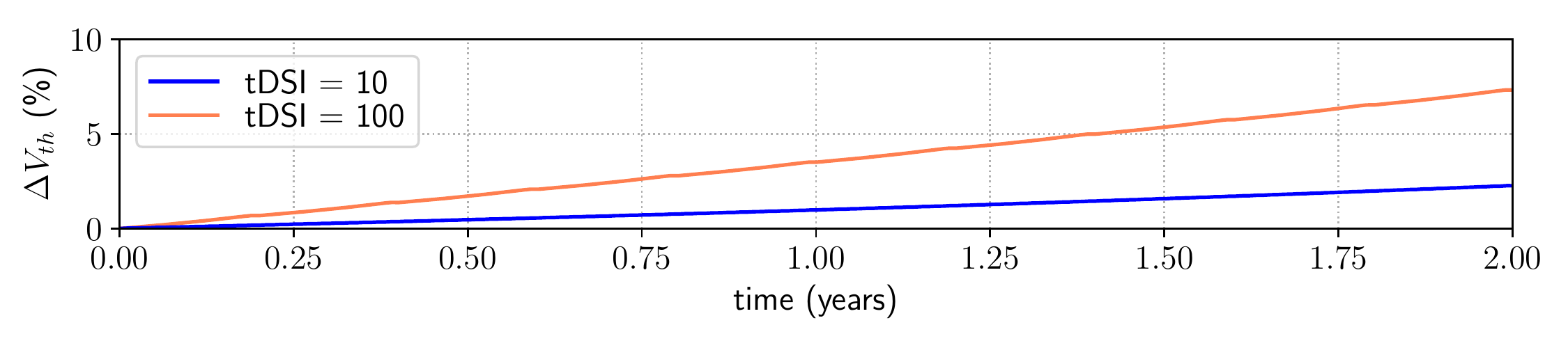}}
 	\vspace{-5pt}
 	\caption{Shift in \ineq{V_\text{th}} for \ineq{tDSI = 10} and \ineq{100}.}
    \vspace{-5pt}
 	\label{fig:bti}
\end{figure}

Figure~\ref{fig:bti} shows the shift in threshold voltage of a transistor in a memory bank's peripheral circuitry when executing a microbenchmark with
\ineq{tDSI} set to \ineq{10} and \ineq{100} requests, respectively.\footnote{{The microbenchmark we use for this evaluation consists of alternating read and write requests to randomly selected PCM locations.}} 

{We observe that the shift in the threshold voltage is higher for larger de-stress interval \ineq{tDSI}. This is because when we set \ineq{tDSI} to a large value, the transistor is exposed to a stress voltage for a long duration between two consecutive de-stress operations. Therefore, the large parameter drift it encounters in this duration cannot be reversed during the next de-stress operation. The parameter drift continues to accumulate over time, resulting in a large shift in the threshold voltage as shown in the figure. Therefore, it is better to set the \ineq{tDSI} to a small value, which allows most of the parameter drifts to be reversed during each de-stress operating, resulting in a lower threshold voltage shift.}

{However, setting a lower \ineq{tDSI} leads to a higher de-stress related performance overhead, which we formulate in Equation~\ref{eq:destress_overhead}. \tech{} exploits this performance-reliability trade-off using its new intelligent memory request scheduler (see Sec.~\ref{sec:access_scheduler}).}

\section{New Aging Model of \tech{}}
\label{sec:aging_formulation}
In this section, we introduce the new aging model of \tech{}. The BTI lifetime~\cite{song2020case,song2020improving,balaji2019framework,mneme,das2014temperature,das2014combined,das2014reinforcement,das2014energy} of a transistor is
\begin{equation}
    \label{eq:MTTF_NBTI}
    \footnotesize \text{MTTF}_\text{BTI} = \frac{A}{V^\gamma}e^{\frac{E_a}{KT}},
\end{equation}
where \ineq{A} and \ineq{\gamma} are material-related constants, \ineq{E_a} is the activation energy, \ineq{K} is the Boltzmann constant, \ineq{T} is the temperature, and \ineq{V} is the overdrive gate voltage of the transistor.\footnote{{Overdrive voltage is defined as the voltage between transistor gate and source ($V_{GS}$) in excess of the threshold voltage ($V_\text{th}$), where $V_\text{th}$ is the minimum voltage required between gate and source to turn the transistor on.}}
BTI failures can also be modeled using the Weibull distribution with a scale parameter \ineq{\alpha} and a slope parameter \ineq{\beta}. {The reliability, defined as the probability of correct operation of the transistor}, at time \ineq{t} is given by~\cite{das2015reliability,huang2011task,bolchini2014lightweight,SriniISCA04}
\begin{equation}
\label{eq:eq1}
\scriptsize R(t) = e^{-\left(\frac{t}{\alpha(V)}\right)^\beta},
\end{equation}
with the corresponding MTTF computed as
\begin{equation}
\label{eq:2}
\scriptsize MTTF = \int_{0}^{\infty}R(t)dt = \alpha(V)\Gamma\left(1+ \frac{1}{\beta}\right),
\end{equation}
where \ineq{\Gamma} is the Gamma function.
Using the expressions for MTTF from Equations~\ref{eq:MTTF_NBTI} and \ref{eq:eq1}, and rearranging, we obtain the expression for the scale parameter \ineq{\alpha} as
\begin{equation}
\label{eq:eq3}
\scriptsize \alpha(V) = {\frac{A}{V^\gamma}e^{\frac{E_a}{KT}}} \bigg{/} {\Gamma\left(1+\frac{1}{\beta}\right)}.
\end{equation}

\begin{figure}[h!]
 	\centering
 	\centerline{\includegraphics[width=0.99\columnwidth]{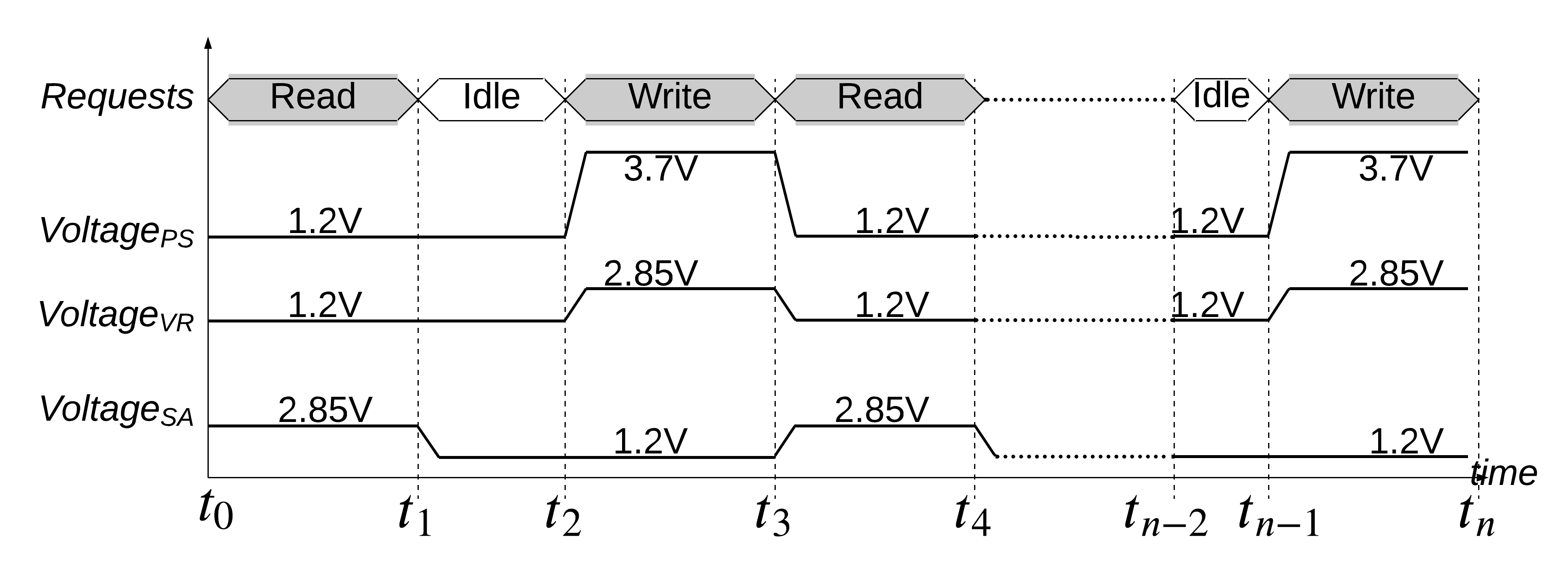}}
 	\caption{Operating voltage of PS, VR, and SA logic blocks in peripheral circuitry of a bank when serving read and write requests. (Overdrive voltage = operating voltage - \ineq{V_\text{th}}).}
 	\label{fig:aging_calculation}
\end{figure}

Figure~\ref{fig:aging_calculation} shows the operating voltage of PS, VR, and SA blocks in the peripheral circuitry of a memory bank when serving read and write requests from the bank. We observe that the operating voltage of the logic blocks in a memory bank's peripheral circuit changes over time based on whether the peripheral circuit is idle or serving read or write requests. Existing aging models such as~\cite{SriniISCA04,das2015reliability,das2013aging} assume constant operating voltage for the logic blocks. {Therefore, these models cannot be effectively used to estimate the aging in a memory bank's peripheral circuitry.} 

We illustrate how we formulate aging of each of these three logic blocks in peripheral circuitry, starting with the PS logic. 
Let \ineq{[t_i,t_{i+1})} be the \ineq{(i+1)}th time interval with \ineq{\Delta t_i = t_{i+1} - t_i} and \ineq{V_i} be the gate overdrive voltage in this time interval \ineq{t_i}. 
The reliability of the PS logic at the start of execution is
\begin{equation}
\label{eq:eq4}
\scriptsize R(t)|_{t=t_0} = 1.
\end{equation}
At the end of the first interval (i.e, after servicing the first read request), the reliability of the PS logic is
\begin{equation}
\label{eq:eq5}
\scriptsize R(t_1^-) = e^{-\left(\frac{t_1}{\alpha(V_0)}\right)^\beta}.
\end{equation}
Using the term $\theta$ to represent reliability degradation during this interval $[t_o,t_1)$, the reliability at the beginning of the second interval (i.e., right after the start of the first idle period) is
\begin{equation}
\label{eq:eq7}
\scriptsize R(t_1^+) = e^{-\left(\frac{t_1+\theta}{\alpha(V_1)}\right)^\beta}.
\end{equation}
Due to the continuity of the reliability function, we can equate Equations~\ref{eq:eq5} \& \ref{eq:eq7} to compute $\theta$ as
\begin{equation}
\label{eq:eq8}
\scriptsize \theta = \left(\frac{\alpha(V_1)}{\alpha(V_0)} - 1\right)t_1.
\end{equation}
Substituting Eq.~\ref{eq:eq8} in Eq.~\ref{eq:eq7}, reliability at time $t_2$ is 
\begin{equation}
\label{eq:eq9}
\scriptsize R(t_2) = e^{-\left(\frac{\Delta t_1}{\alpha(V_1)} + \frac{\Delta t_0}{\alpha(V_0)}\right)^{\beta}}.
\end{equation}
We can extend this equation to compute the reliability of the PS logic at the end of execution (i.e., after servicing the last write request from the bank in Fig.~\ref{fig:aging_calculation}) as
\begin{equation}
\label{eq:eq10}
\scriptsize R(t_s) = e^{-\left(\sum_{i=1}^{n}\frac{\Delta t_i}{\alpha(V_i)}\right)^{\beta}},
\end{equation}

The aging \ineq{\mathcal{A}_\text{PS}} of the PS logic is 
\begin{equation}
\label{eq:eq11_pre1}
\scriptsize \mathcal{A}_\text{PS} = \sum_{i=1}^{n}\frac{\Delta t_i}{\alpha(V_i)}, \text{ such that } R(t_s) = e^{-(\mathcal{A}_\text{PS})^{\beta}},
\end{equation}
where the scaling factor \ineq{\alpha(V_i)} can be calculated using Eq. \ref{eq:eq3}.

We observe that Eq.~\ref{eq:eq11_pre1} follows the \textit{associative property}, a direct reflection of the underlying BTI failure mechanism. In other words, the aging accrued in each bank's peripheral circuitry is independent of the order in which the reads and writes are scheduled to the bank. Eq.~\ref{eq:eq11_pre1} can be rewritten using memory timing parameters as
\begin{equation}
    \label{eq:eq11}
    \scriptsize \mathcal{A}_\text{PS} = n_r\cdot \mathcal{U}_r + n_w\cdot \mathcal{U}_w + n_i\cdot\mathcal{U}_i, \text{ where }
\end{equation}
\vspace{-10pt}
\begin{scriptsize}
\begin{align*}
    &\mathcal{U}_r = \frac{tRC_r}{\alpha(1.2)}, \mathcal{U}_w = \frac{tRC_w}{\alpha(3.7)}, \text{ and } \mathcal{U}_i = \frac{1}{\alpha(1.2)}
\end{align*}
\end{scriptsize}
\normalsize
where, \ineq{tRC} is the row cycle time, \ineq{n_r} and \ineq{n_w} are the number of read and write requests, respectively, and \ineq{n_i} is the number of memory clock cycles for which the PS logic is idle. {\ineq{\mathcal{U}_r} and \ineq{\mathcal{U}_w} represent respectively, the aging accrued in peripheral circuitry when serving a read and a write request, and \ineq{\mathcal{U}_i} represents the aging accrued per clock cycle when the peripheral circuitry is idle.
\ineq{\mathcal{U}_r, \mathcal{U}_w}, and \ineq{\mathcal{U}_i} are called \textit{unit aging parameters}, which the memory controller uses to track the aging of the PS logic in peripheral circuitry by simply recording 1) the number of read and write requests that are served from the bank, and 2) the number of idle clock cycles during workload execution.} We note that these factors (read/write requests and the idle periods) cannot be known with certainty at design-time. Therefore, design-time aging estimates are not accurate.

The aging of VR and SA logic blocks 
(represented as \ineq{\mathcal{A}_\text{VR}} and \ineq{\mathcal{A}_\text{SA}}, respectively) can be computed in a similar way using Eq.~\ref{eq:eq11}.
To obtain the overall aging, we combine these individual aging values considering the peripheral circuitry to be a \textit{series} failure system, where the first instance of any logic block failing causes the entire peripheral circuit to fail. Therefore, the overall aging  is
\begin{equation}
    \label{eq:overall_aging}
    \footnotesize \mathcal{A} = \max\{\mathcal{A}_\text{PS},\mathcal{A}_\text{VR},\mathcal{A}_\text{SA}\}.
\end{equation}

\section{Decoupled Peripheral Circuit Architecture of \tech{}}\label{sec:architectural_change}
\begin{figure}[h!]
 	\centering
    \vspace{-10pt}
 	\centerline{\includegraphics[width=0.99\columnwidth]{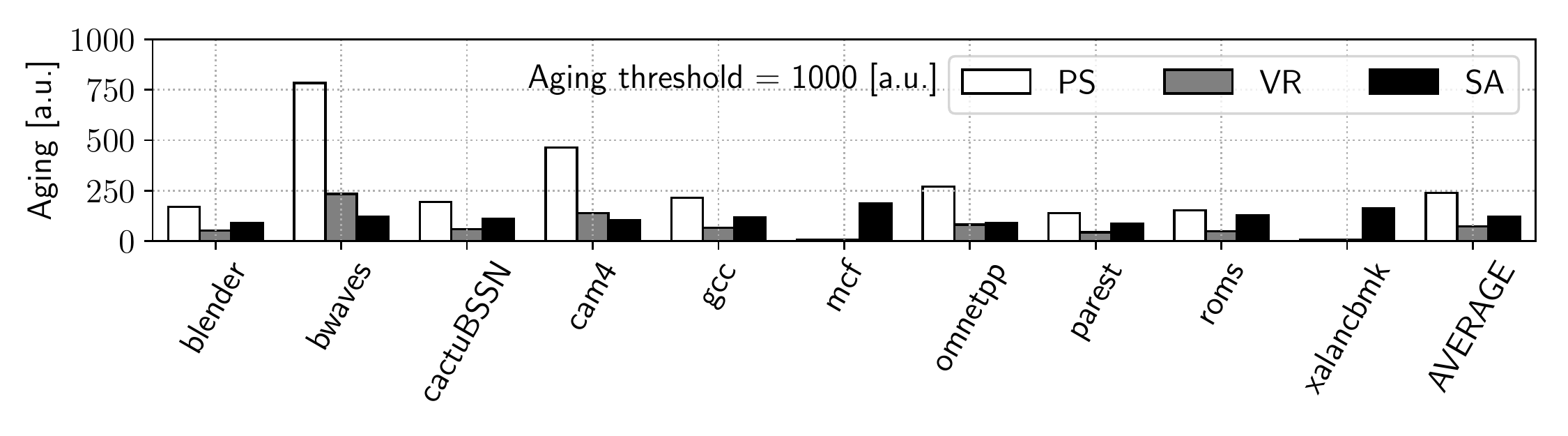}}
 	\vspace{-10pt}
 	\caption{Average aging in arbitrary units (a.u.) of the PS, VR, and SA logic blocks in peripheral circuitry, at the time when they are de-stressed for our evaluated workloads.}
    \vspace{-10pt}
 	\label{fig:laser_aging}
\end{figure}

In the baseline system, when  peripheral circuitry is de-stressed, its three logic blocks (PS, VR, and SA) are de-stressed simultaneously. Once de-stressed, these logic blocks take several memory cycles (\ineq{tDSC}) before they can be used to serve memory requests again. In recent designs, \ineq{tDSC = 10} cycles~\cite{pcm_book}. Therefore, frequent de-stress operations can lead to high performance overhead (Eq.~\ref{eq:destress_overhead}). To reduce this overhead, we analyze the average aging of the PS, VR, and SA blocks in a memory bank's peripheral circuitry at the time when they are de-stressed during workload execution.
Figure~\ref{fig:laser_aging} plots these results for the workloads described in Section~\ref{sec:evaluation} 
with \ineq{tDSI} set to \ineq{100}.
We make the following two key observations.

First, the average aging of these logic blocks varies widely across different workloads due to the difference in memory requests.
Second, at the time when a de-stress is initiated, the average aging for these three logic blocks are different and lower than the aging threshold.
This is because in the baseline design, a peripheral circuit is de-stressed at its entirety, when the aging in any one of its three logic blocks exceeds the aging threshold.

Based on these observations and the connectivity of the logic blocks to the two charge pumps (see Fig.~\ref{fig:peripheral_circuit}), we introduce an isolation transistor (\texttt{M}) to decouple the VR logic from the PS logic inside the write driver, allowing us to track and de-stress the logic blocks individually, as opposed to de-stressing the entire peripheral circuitry at once. Table~\ref{tab:aging_control} summarizes the new controls, which we enable using the isolation transistor.


Using these new decoupled control mechanism, \tech{}'s request scheduler (Section~\ref{sec:access_scheduler}) can de-stress logic blocks in a bank's peripheral circuitry off the critical path of accesses, lowering bank occupancy and improving performance.
{We observe that the read charge pump is shared between the SA and VR logic blocks (see Figure~\ref{fig:peripheral_circuit}). Therefore, when \tech{} de-stresses the SA because SA's aging exceeds the critical aging threshold, the VR logic also gets de-stressed, preventing the write driver from serving write requests. To address this, we exploit the decoupled program and verify-based write operations in PCM~\cite{frulio2016adaptive}. If a write request needs to be scheduled concurrently with the de-stress operation of SA, \tech{} schedules only the program step of the write operation (which utilizes the PS block) concurrently with the de-stress operation, while the verify step is scheduled after the de-stress operation completes.}

\vspace{-5pt}
\begin{table}[h!]
\setlength{\tabcolsep}{5pt}
\renewcommand{\arraystretch}{1.0}
\centering
{\fontsize{8}{10}\selectfont
\begin{tabu}{c c | c c c}
    \tabucline[2pt]{-}
    \multicolumn{2}{c|}{\textbf{Charge Pump Control}} & \multicolumn{3}{c}{\textbf{Peripheral Circuit Action}}\\ \hline
    \textbf{Read} & \textbf{Write} & \textbf{PS} & \textbf{VR} & \textbf{SA}\\
    \hline
    \hline
    \multicolumn{5}{c}{\textcolor{blue}{Baseline Control}}\\
    \hline
    Active & Active & Active & Active & Active\\
    Discharged & Discharged & De-stress & De-stress & De-stress\\
    \hline
    \multicolumn{5}{c}{\textcolor{blue}{Proposed Decoupled Control}}\\
    \hline
    Active & Active & Active & Active & Active\\
    Discharged & Active & Active & De-stress & De-stress\\
    Active & Discharged & De-stress & Active & Active\\
    Discharged & Discharged & De-stress & De-stress & De-stress\\
    \tabucline[2pt]{-}
\end{tabu}
}
\vspace{5pt}
\caption{Controlling de-stress ops. using charge pumps.}
\label{tab:aging_control}
\end{table}
\vspace{-25pt}


\section{Intelligent Memory Request Scheduler of \tech{}}
\label{sec:access_scheduler}

{We design a new memory request scheduling policy to control the aging in the peripheral circuitry within each memory bank using our new aging model (Section~\ref{sec:aging_formulation}) and the decoupled peripheral circuit architecture (Section~\ref{sec:architectural_change}).}

\subsection{High-level Overview}

\begin{figure}[h!]
 	\centering
    \vspace{-10pt}
 	\centerline{\includegraphics[width=0.85\columnwidth]{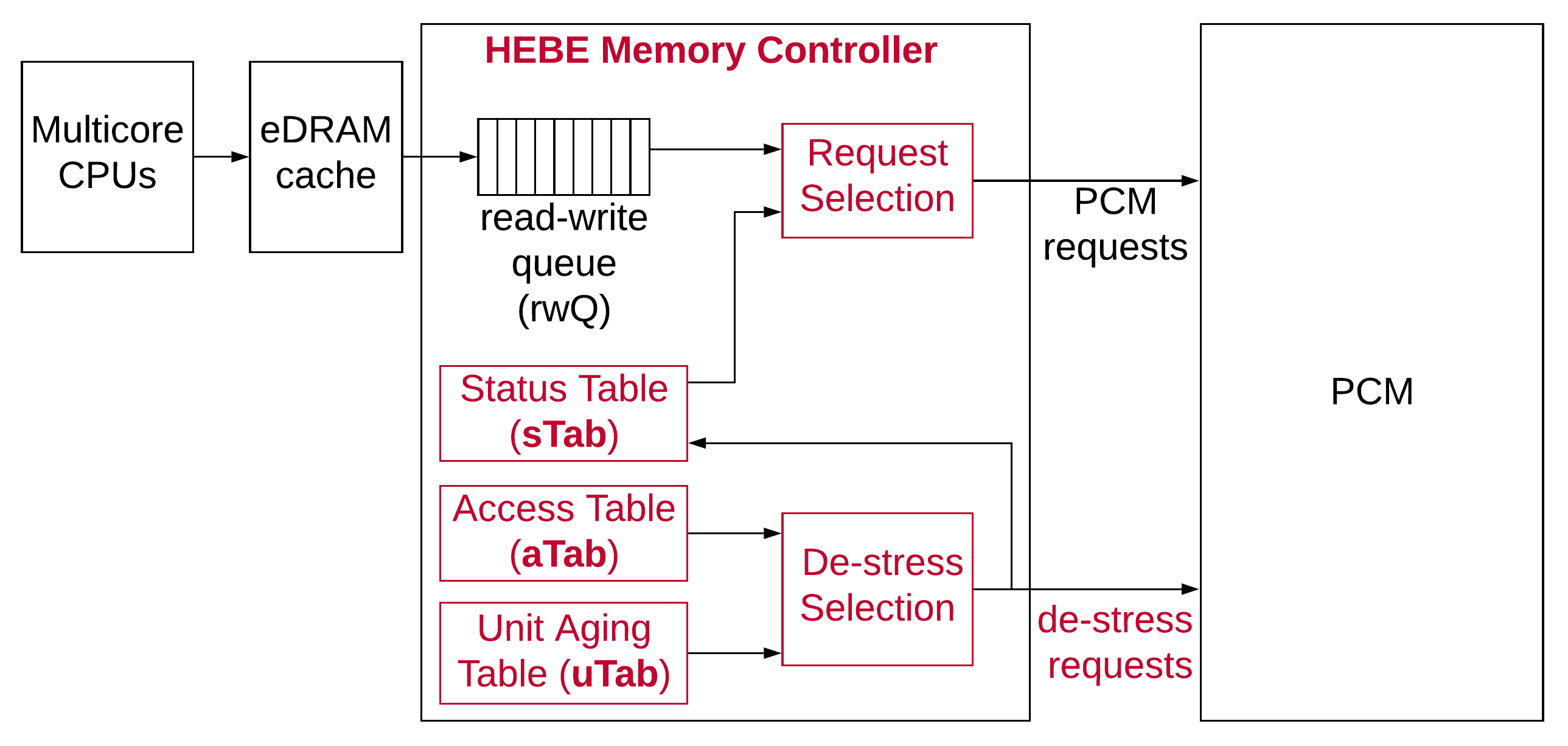}}
 	\vspace{-10pt}
 	\caption{Request and de-stress scheduling in \tech{}.}
    \vspace{-15pt}
 	\label{fig:hebe}
\end{figure}

We describe \tech{} in the context of DRAM-PCM hybrid memory, where embedded DRAM (eDRAM) is used as a write cache to PCM main memory as shown in Figure~\ref{fig:hebe}.\footnote{Even though we use eDRAM as cache to PCM in our implementation and evaluations, \tech{} is applicable to any type of hybrid memory or standalone PCM memory.} The baseline memory controller architecture consists of a read-write queue (rwQ) to buffer PCM requests.
The key idea of \tech{}'s scheduling policy is to 1) improve performance by lowering the de-stress overhead, 
2) minimize 
wasted memory cycles during which a bank is idle with its peripheral circuitry accruing BTI aging, and 
%
3) prevent a request from being delayed too much.

\subsection{Detailed Design of \tech{}}
Figure \ref{fig:hebe} shows the detailed design of \tech{}, which introduces five new components to the baseline memory controller design as highlighted in the figure.

The \textit{first} component is the \emph{status table} (\textbf{sTab}). \tech{} uses this table to record if a memory bank is available to serve a PCM request. sTab requires one 1-bit entry for each PCM bank. For 128 banks in a 128GB PCM (see our simulation parameters in Table~\ref{tab:simulation_parameters}), \tech{} requires 128 bits of storage for sTab.

The \textit{second} component is the \emph{access table} (\textbf{aTab}). \tech{} uses this table to record the number of memory cycles for which a memory bank's peripheral circuitry is active since the last de-stress operation of the bank. Since peripheral circuitry operates at a different voltage when it is idle than when it is serving a read or a write request, the aging model of \tech{} requires the exact number of cycles for which a peripheral circuit is idle and serving read and write requests. Therefore, each aTab entry contains one 16-bit field for recording the idle cycles, and two 4-bit fields for recording the number of read and write requests. For 128GB PCM with 1GB per bank, \tech{} requires 3Kb (= 128 x 24 bits) of storage.

The \textit{third} component is the \textit{unit aging table} (\textbf{uTab}). {\tech{} uses this table to store the unit aging parameters (Eq.~\ref{eq:eq11}). Since the three unit aging parameters are the same for every peripheral circuitry in PCM, there are only three 32-bit entries in this table, one for each of these parameters, requiring a total of 96 bits for uTab.}\footnote{{For simplicity, we have not considered process variation across different peripheral circuitry of different PCM banks.}}

\begin{figure}[h!]
 	\centering
    \vspace{-10pt}
 	\centerline{\includegraphics[width=0.75\columnwidth]{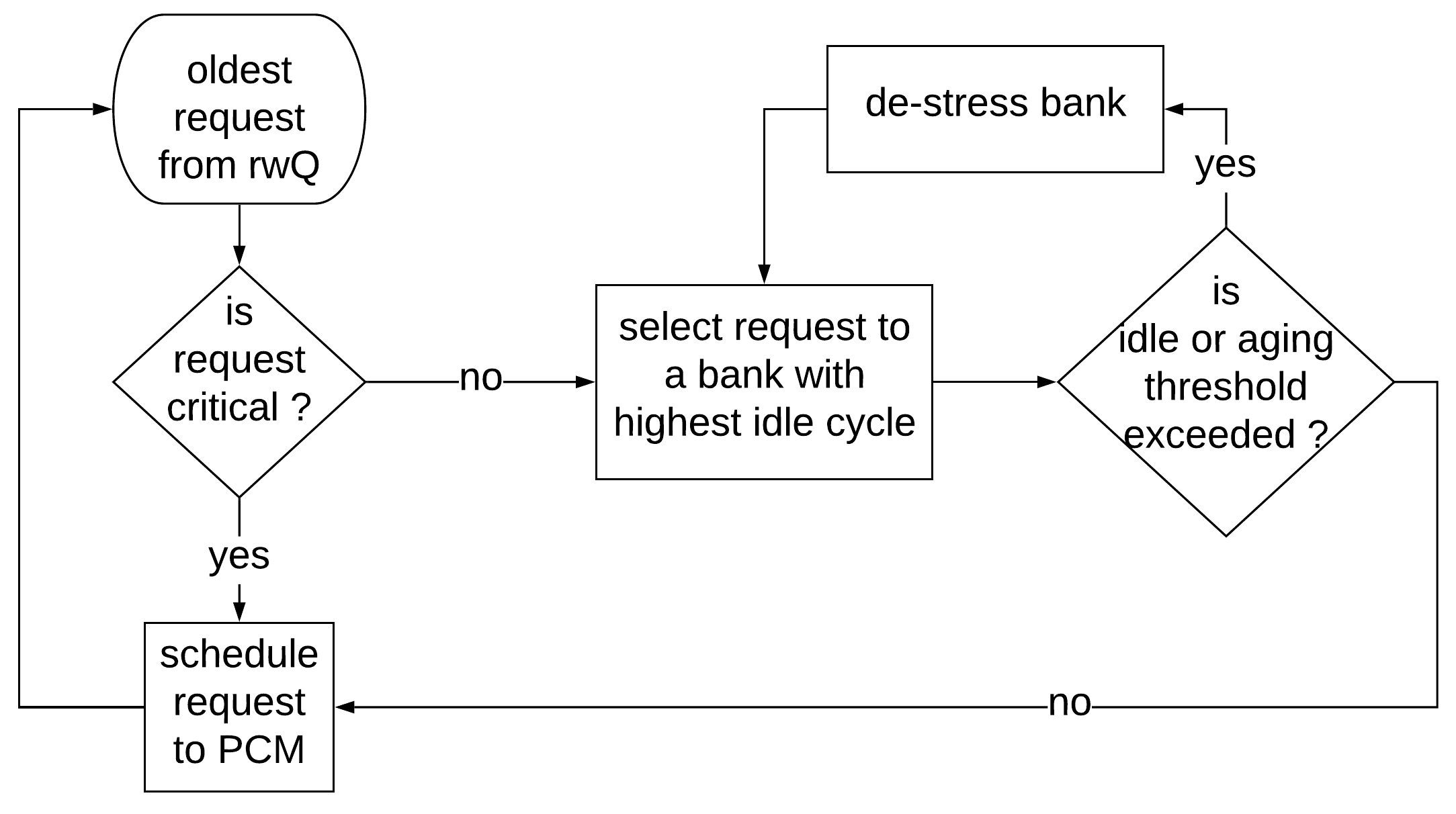}}
 	\vspace{-10pt}
 	\caption{Memory request and de-stress selection in \tech{}.}
    \vspace{-10pt}
 	\label{fig:flowchart}
\end{figure}

The \textit{fourth} component is the \textit{request selection}. \tech{} uses this component to select a request from the rwQ to schedule to PCM. Figure~\ref{fig:flowchart} shows the flowchart of \tech{}'s request selection mechanism. 
After scheduling a request from the rwQ, the memory controller checks to see if the number of clock cycles for which a request is outstanding in the rwQ is smaller than the \textit{backlogging threshold} (\ineq{th_b}). If the backlogging threshold is exceeded, the request is de-queued and served next. 
{Otherwise, 
the memory controller selects an outstanding request from the rwQ that is to a bank whose peripheral circuitry has the highest number of idle cycles since the time it served a request from the bank.}

{The \textit{final} component is the \textit{de-stress selection logic}. \tech{} uses this component to schedule de-stress operations in PCM banks.
For this purpose, \tech{} uses two thresholds -- the aging threshold (\ineq{th_a}) and the idle threshold (\ineq{th_i}). The aging threshold is used to control the aging of peripheral circuitry in PCM in order to achieve a target lifetime. The idle threshold is used to limit the duration during which a peripheral circuit accrues aging without doing any useful work.
The de-stress selection logic is also shown in Figure~\ref{fig:flowchart}.
If the selected memory request is to a bank whose peripheral circuitry exceeds either of the two thresholds, the memory controller schedules a de-stress operation to the bank. Otherwise, the request is scheduled to PCM.
}


\subsection{Overhead of \tech{}}
\tech{} requires a total storage of 3.2Kb for a PCM memory of 128GB capacity and 128 banks. The timing overhead of the request and de-stress selection is overlapped with the timing of an ongoing read or a write request, incurring minimal impact on the critical path of PCM read and write accesses. Therefore, \tech{}'s request scheduling introduces marginal performance overhead. On the contrary, \tech{} improves performance compared to other approaches by reducing the de-stress related performance bottleneck (see Section~\ref{sec:res_system_performance}).

\section{Evaluation Methodology}
\label{sec:evaluation}
We evaluate \tech{} for phase-change memory (PCM), one of the matured NVM technologies. We configure PCM as main memory with eDRAM as its write cache. This is similar to the architecture of IBM POWER 9~\cite{sadasivam2017ibm}.
Our simulation framework includes
the following components with parameters listed in Table~\ref{tab:simulation_parameters}.
\begin{itemize}
	\item Cycle-level in-house x86 multi-core simulator. We configure this to simulate 8 out-of-order cores. 
	\item Main memory simulator, closely matching the JEDEC 
	Nonvolatile Dual In-line Memory Module (NVDIMM) specifications~\cite{jedecnvdimm2017}. This simulator is composed of Ramulator~\cite{kim2016ramulator}, to simulate DRAM and an cycle-level in-house NVM simulator, based on NVMain~\cite{poremba2015nvmain}.
	\item Power and latency for DRAM and NVM are based on Intel/Micron's 3D Xpoint specification~\cite{pcm_book}. Energy is modeled for DRAM using DRAMPower~\cite{chandrasekar2012drampower} and for NVM using NVMain with parameters from~\cite{pcm_book}.
\end{itemize}

\begin{table}[h!]
\setlength{\tabcolsep}{5pt}
\renewcommand{\arraystretch}{1.0}
\centering
{\fontsize{8}{10}\selectfont
\begin{tabu}{r p{5.5cm}}
    \tabucline[2pt]{-}
    Processor & 8 cores, 3 GHz, out-of-order\\
    L1-I/D cache & Private 64KB per core, 4-way\\
    L2 cache & Shared, 4MB, 8-way \\
    \hline
    DRAM & 8GB, Micron DDR3\\
    Main Memory & 1 channel, 8 rank/channel, 8 banks/rank, 128 sub-arrays/bank, 512 rows/sub-array\\
    \hline
    PCM & 128GB, Micron DDR3 \cite{pcm_book}\\
    Main Memory & 2 channels, 1 rank/channel, 8 banks/rank, 64 partitions/bank, 128 tiles/partition, 4096 rows/tile\\
    \tabucline[2pt]{-}
\end{tabu}
}
\vspace{5pt}
\caption{Major simulation parameters.}
\label{tab:simulation_parameters}
\end{table}
\vspace{-20pt}

Table~\ref{tab:min_max_latency} reports the timing parameters for PCM reads and writes. These parameters are based on Micron's 128GB PCM module~\cite{pcm_book}. 

\begin{table}[h!]
\setlength{\tabcolsep}{5pt}
\renewcommand{\arraystretch}{1.3}
\centering
{\fontsize{8}{10}\selectfont
\begin{tabu}{c c c c c c}
    \tabucline[2pt]{-}
    & \textbf{tRCD} & \textbf{tRAS} & \textbf{tRP} & \textbf{tRC} & \\\cline{2-6}
    Read & 3.75ns & 55.25ns & 1ns & 56.25ns &  \\
    \hline
    & \textbf{tRCD} & \textbf{tBURST} & \textbf{tWR} & \textbf{tRP} & \textbf{tRC}\\\cline{2-6}
     Write & 75ns & 15ns & 190ns & 1ns & 209.75ns \\
    \tabucline[2pt]{-}
\end{tabu}
}
\vspace{5pt}
\caption{{PCM timing parameters based on \cite{pcm_book}.}}
\label{tab:min_max_latency}
\end{table}
\vspace{-20pt}

We evaluate 10 billion instructions of ten
workloads from the SPEC CPU2017 benchmarks~\cite{bucek2018spec} (see Table~\ref{tab:spec}).






We evaluate the following techniques.
\begin{itemize}
    \item \textit{\underline{Baseline}}~\cite{datacon} de-stresses peripheral circuitry of each NVM bank with a fixed \ineq{tDSI} of 100, without tracking their aging. Memory requests are scheduled using the FR-FCFS policy~\cite{RixnerISCA2000,zuravleff1997controller}. 
    \item \textit{\underline{\tech{}}} tracks the aging in CMOS transistors in peripheral circuitry of each bank and de-stresses them only when their aging exceeds the aging threshold. A peripheral circuitry is de-stressed based on the maximum aging of its logic blocks.
    \item \textit{\underline{{Decoupled-\tech{}}}} is based on \tech{}. Each peripheral circuitry is decoupled to de-stress its logic blocks independently.
\end{itemize}

\vspace{-5pt}
\begin{table}[h!]
\setlength{\tabcolsep}{5pt}
\renewcommand{\arraystretch}{1.3}
\centering
{\fontsize{8}{10}\selectfont
\begin{tabu}{r p{6cm}}
    \tabucline[2pt]{-}
    single-core & 8 copies each of blender, bwaves, cactuBSSN, cam4, gcc, mcf, omnetpp, parset, roms, xalancbmk\\
    \tabucline[2pt]{-}
\end{tabu}
}
\vspace{5pt}
\caption{Evaluated workloads.}
\label{tab:spec}
\end{table}
\vspace{-15pt}

\subsection{Aging Parameters}
To compute aging, the slope parameter of Weibull distribution is set to \ineq{\beta = 2}, and the operating temperature is set to \ineq{300K}. Other fitting parameters are adjusted to achieve an MTTF of {2 years} in the baseline system, corresponding to a threshold voltage shift of 10\%. This is what is typically accepted as the maximum allowed \ineq{V_\text{th}} degradation before timing errors begin to appear~\cite{balaji2019framework,song2020case,song2020improving,twisha_thermal,twisha_endurance,das2012fault,das2013energy,das2012faultRSP,das2015reliability,das2013aging}.

\section{Results and Analyses}
\label{sec:results}
\subsection{Overall System Performance}
\label{sec:res_system_performance}
\begin{figure}[h!]
	\centering
	\vspace{-10pt}
	\centerline{\includegraphics[width=0.99\columnwidth]{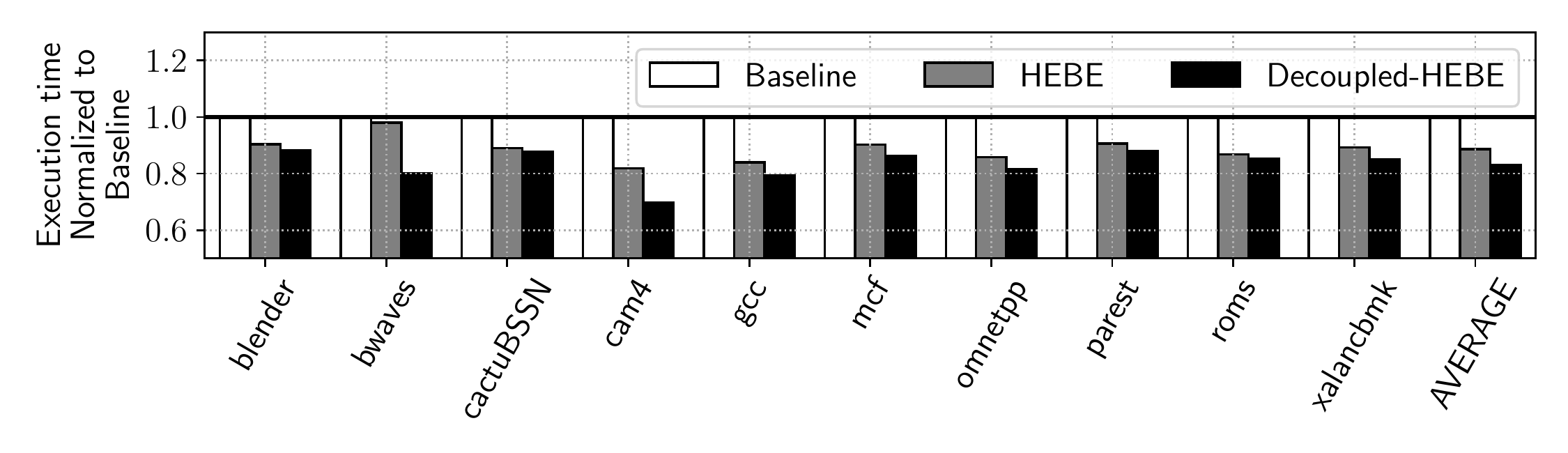}}
	\vspace{-15pt}
	\caption{Execution time, normalized to Baseline.}
	\vspace{-10pt}
	\label{fig:perf}
\end{figure}

Figure~\ref{fig:perf} plots the execution time of each workload for the evaluated systems normalized to Baseline. We make the following two key observations.

First, the average execution time of \tech{} is 12\% lower than Baseline. This improvement is because \tech{} has lower de-stress overhead than Baseline due to \tech{}'s dynamic policy to opportunistically de-stress each peripheral circuitry \textit{only when} its aging exceeds the aging threshold.
Baseline, on the other hand, uses a fixed de-stress interval of 100 without tracking the exact aging.
Second, the average execution time of Decoupled-\tech{} is 6\% lower than \tech{}. This improvement is because Decoupled-\tech{} de-stresses the logic blocks in a memory bank's peripheral circuits off the critical path of read and write accesses from the bank, reducing bank occupancy and improving performance.

\subsection{Overall MTTF}
\label{sec:res_lifetime_reliability}
\begin{figure}[h!]
	\centering
	\vspace{-10pt}
	\centerline{\includegraphics[width=0.99\columnwidth]{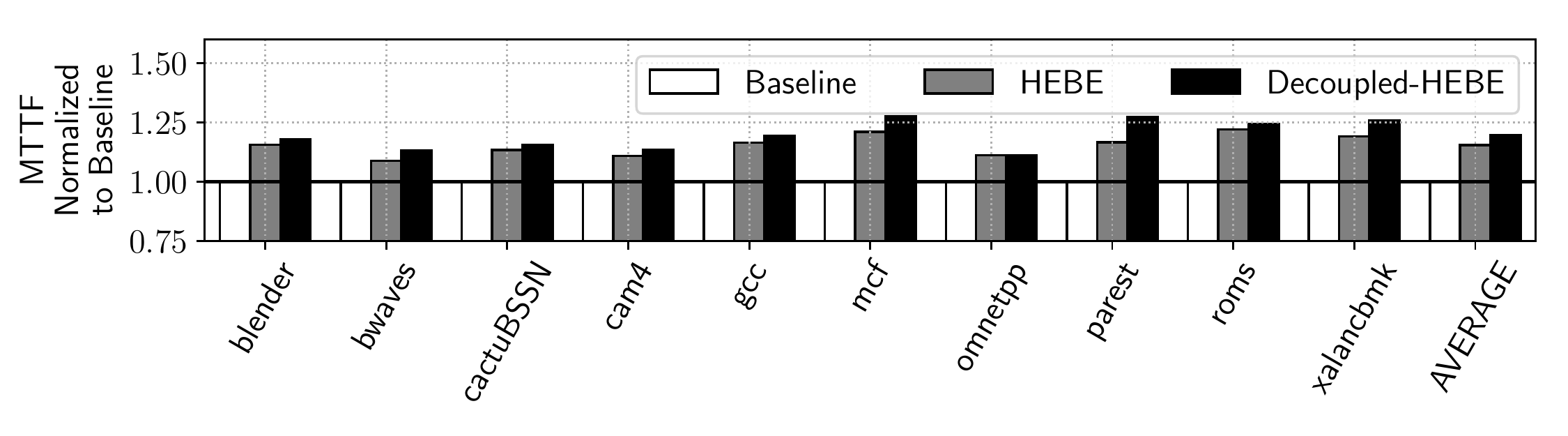}}
	\vspace{-15pt}
	\caption{MTTF, normalized to Baseline.}
	\vspace{-10pt}
	\label{fig:aging}
\end{figure}

Figure~\ref{fig:aging} plots the MTTF of each  workload for the evaluated systems normalized to Baseline. We make the following two key observations.

First, the average MTTF of \tech{} is 16\% higher than Baseline. This improvement is because 1) \tech{} does not allow the aging of any peripheral circuitry in PCM to exceed the aging threshold and 2) the aging-aware access scheduling policy of \tech{} minimizes the number of wasted memory cycles for which a peripheral circuitry accrues aging while being idle. Second, the average MTTF of Decoupled-\tech{} is 3.4\% higher than \tech{}. 
{
This improvement is because 
\tech{} needs to wait for an ongoing PCM read or write request to complete before it can schedule the de-stress operation of a peripheral circuitry. 
Therefore, the circuitry continues to age before it is eventually de-stressed, lowering its MTTF. 
On the other hand, Decoupled-\tech{} can schedule the de-stress operation of a logic block in a bank's peripheral circuitry independently and in parallel to ongoing read and write requests to the bank. Therefore, the MTTF of Decoupled-\tech{} is higher than \tech{}.
}


\subsection{De-stress Overhead}
\label{sec:res_deso}
\begin{figure}[h!]
	\centering
	\centerline{\includegraphics[width=0.99\columnwidth]{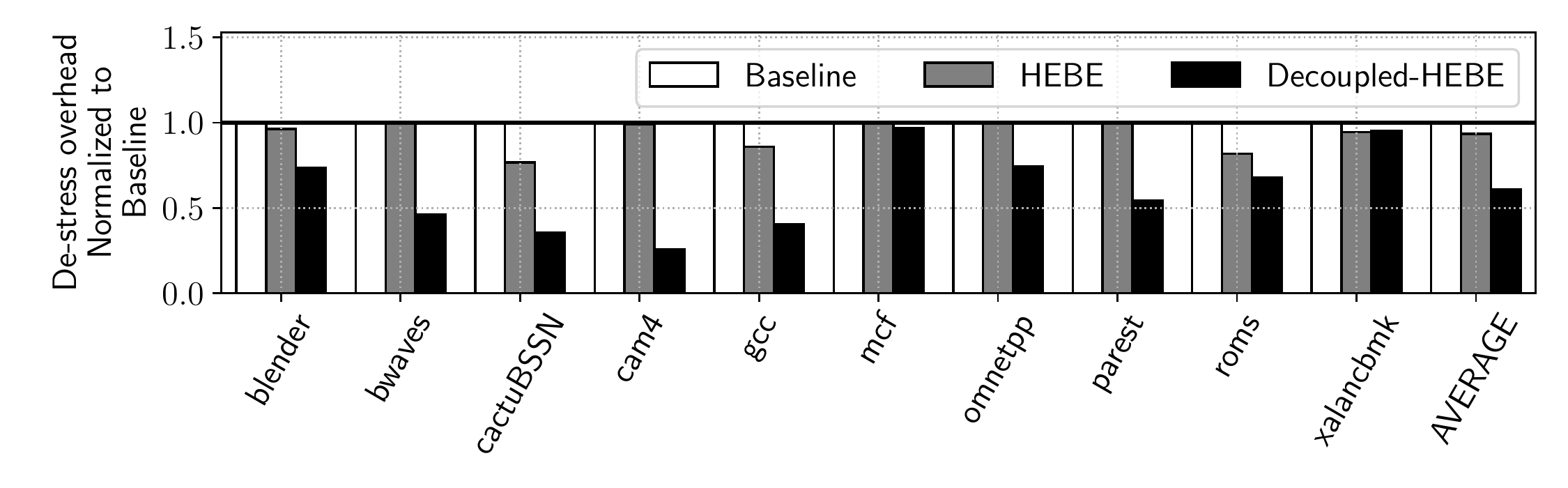}}
	\vspace{-15pt}
	\caption{De-stress overhead, normalized to Baseline.}
	\vspace{-10pt}
	\label{fig:deso}
\end{figure}

Figure~\ref{fig:deso} plots the de-stress overhead (Eq.~\ref{eq:destress_overhead}) of each workload for each evaluated system normalized to Baseline. We make the following two key observations.

First, the average de-stress overhead of \tech{} is 6.6\% lower than Baseline. This improvement is because \tech{} increases the de-stress interval (\ineq{tDSI}) by accurately tracking the aging of each peripheral circuitry dynamically, de-stressing it \textit{only when} aging exceeds a threshold. 
Baseline uses a fixed \ineq{tDSI} of 100.
Second, the average de-stress overhead of Decoupled-\tech{} is  35\% lower than \tech{}. This improvement is due to the reduction of the de-stress cycle time (\ineq{tDSC}), which is achieved by de-stressing the logic blocks in a bank's peripheral circuitry independently and in parallel to ongoing read and write requests to the bank.

\subsection{Effect of Aging Threshold}\label{sec:res_threshold}
\begin{figure}[h!]
	\centering
	\vspace{-10pt}
	\centerline{\includegraphics[width=0.99\columnwidth]{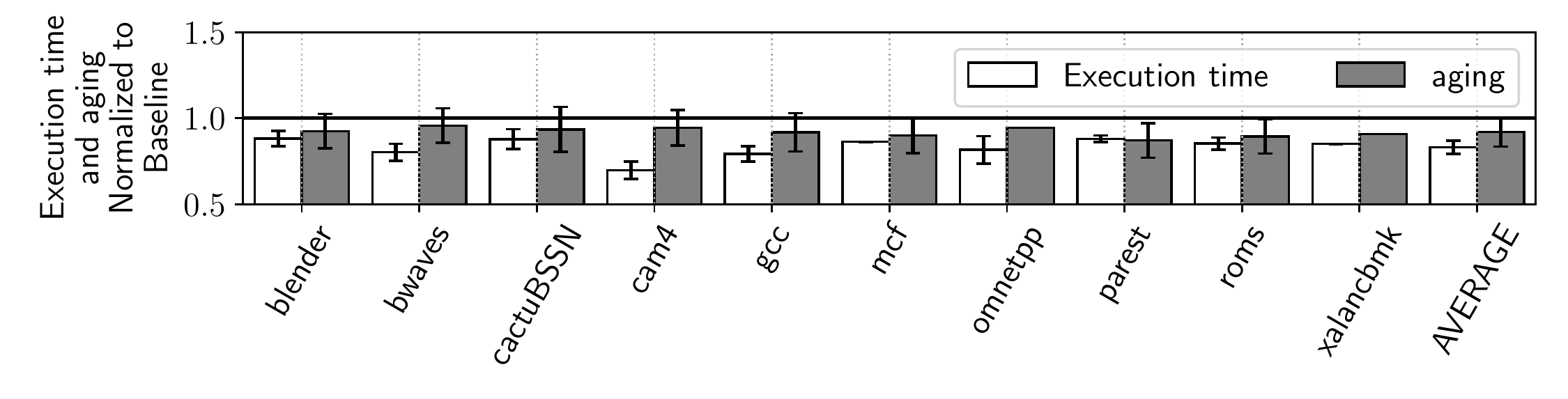}}
	\vspace{-10pt}
	\caption{Execution time and aging of \tech{}, normalized to Baseline, as a function of the aging threshold.}
	\vspace{-10pt}
	\label{fig:bar_age_thr}
\end{figure}

Figure~\ref{fig:bar_age_thr} reports the execution time and aging of each of our workloads using \tech{}, normalized to Baseline. The height of a bar represents \tech{}'s result with the default aging threshold of 1000 units. An error bar represents the variation obtained by changing the aging threshold from 500 units to 2000 units. We make the following observation. 

When we set a stricter aging threshold (e.g., 500 units), execution time increases and aging decreases. This is because when the aging threshold is lowered, the high-voltage exposure time of the peripheral circuitry in each memory bank reduces, reducing the accrued aging. However, performance degrades due to the high de-stress overhead (see Eq.~\ref{eq:destress_overhead}). Conversely, when we relax the aging threshold (e.g., 2000 units), the de-stress interval increases, reducing the de-stress overhead and increasing performance. However, aging is now higher because of longer exposure to high-voltage stress.


\subsection{Temperature Dependency}
\begin{figure}[h!]
	\centering
	\vspace{-10pt}
	\centerline{\includegraphics[width=0.99\columnwidth]{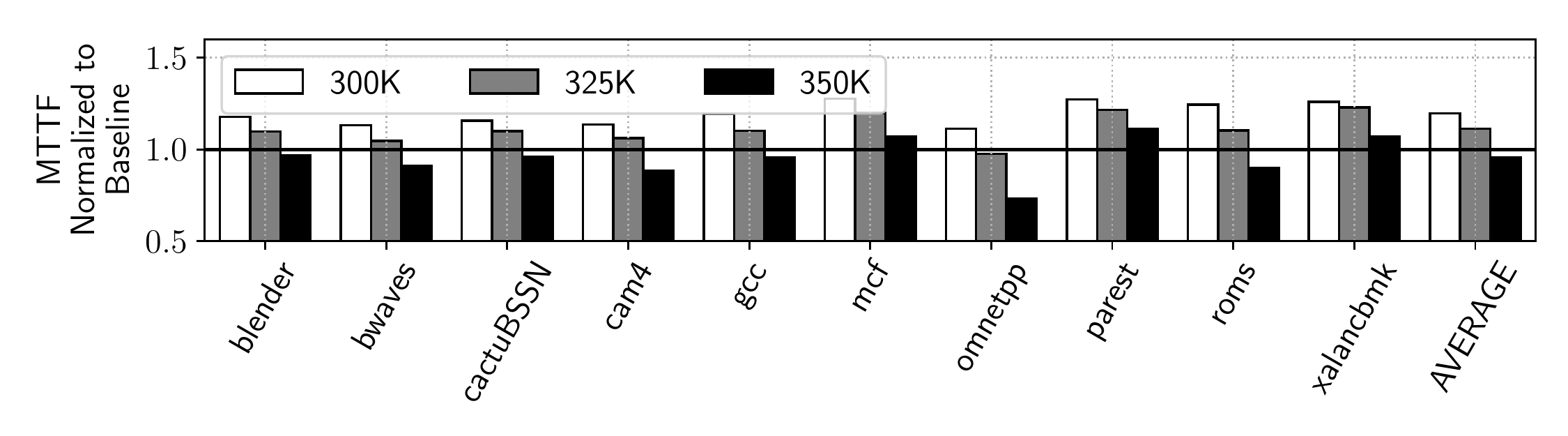}}
	\vspace{-15pt}
	\caption{MTTF at 325K and 350K normalized to Baseline.}
	\label{fig:temperature}
\end{figure}

Figure~\ref{fig:temperature} plots MTTF of Decoupled-\tech{} at 300K, 325K, and 350K normalized to Baseline at 300K for each evaluated application. 
We observe that MTTF decreases with increase in temperature. MTTF at 325K and 350K are higher than at 300K by an average of 7\% and 26\%, respectively.
These results follow directly from our aging formulation, which incorporates temperature using the scaling parameter \ineq{\alpha} in Eq. \ref{eq:eq3}. This parameter grows exponentially with temperature, resulting in a corresponding exponential increase in aging. 
More aging leads to larger shift in threshold voltage.

\section{Related Works}
\label{sec:related_works}
To our knowledge, this is the first work that exploits a workload's access characteristics to \emph{dynamically} control the length of the de-stress interval of peripheral circuitry in each memory bank, improving both performance and MTTF of PCM-based main memory. Many works propose optimizations for PCM. Recent examples include architecture optimization~\cite{QureshiISCA09,LeeISCA2009,yavits2020wolfram}, performance and energy optimization~\cite{yoon2012row,yoon2014efficient,palp,datacon,kim2019ll,arjomand2016boosting}, wear leveling~\cite{mneme,zhang2017toss,zhang2018retrofit}, and memory controller optimizations~\cite{zhao2014firm}. See~\cite{xia2015survey} for a survey of these and other similar approaches. \tech{} can be combined with most of these techniques.

Many works propose memory latency reduction, refresh optimization, energy reduction, and request scheduling methods to enhance system performance, fairness, quality of service, or security~\cite{ren2015thynvm,seshadri2015gather,subramanian2014blacklisting,subramanian2016bliss,chang2017understanding,khan2016parbor,chang2016understanding,liu2012raidr,usui2016dash,ausavarungnirun2012staged,RixnerISCA2000,chang2014improving,kim2010atlas,mutlu2008parallelism,kim2010thread,meza2012enabling,lu2014loose,pelley2014memory,ipek2008self,hassan2016chargecache,lee2015adaptive,qureshi2015avatar,patel2017reach,deng2011memscale,david2011memory,kim2018dram,kim2018solar,kim2019d,mutlu2007stall,nesbit2006fair,rixner2004memory}. None of these works consider aging of phase change memory in their scheduling decisions. Our aging-aware scheduling mechanism can be incorporated into other memory controller designs that aim to improve other metrics.


\section{Conclusions}
\label{sec:conclusions}
We introduce \tech{}, a new mechanism that can dynamically track and control the aging of transistors in peripheral circuitry of each memory bank, improving both performance and aging of NVM-based main memory.
\tech{} is built on three {novel} contributions.
First, we propose a {new, accurate} analytical model to {dynamically} track aging in response to the memory controller's request scheduling decisions.
Second, we develop a {new, intelligent} request scheduler that exploits this aging model at {run time} 
to decide {when} peripheral circuitry in NVM  must be de-stressed.
Third, we 
decouple logic blocks in peripheral circuitry operating at different voltages, allowing these blocks to be de-stressed independently 
and
off the 
critical path of execution, improving performance.
We evaluate \tech{} for DRAM-NVM hybrid main memory and show the significant performance and MTTF improvement.
We \textbf{conclude} that \tech{} is a \emph{simple yet powerful} mechanism to dynamically manage the aging in non-volatile main memory, and improve both performance and lifetime via its intelligent request scheduling decisions.


\section*{Acknowledgments}
This work is supported by the National Science Foundation Faculty Early Career Development Award CCF-1942697 (CAREER: Facilitating Dependable Neuromorphic Computing: Vision, Architecture, and Impact on Programmability).

\balance
\bibliographystyle{IEEEtranSN}
\bibliography{commands,disco,external}


\end{document}